\documentclass[final] {aipproc} 
\usepackage{psfrag}
\usepackage{graph icx}
\layoutstyle{6x9}

\begin{document}

\title{Causal and homogeneous networks}


\author{P. Bia\l{}as}{
address={M.Smoluchowski Institute of Physics, Jagellonian University,
         Reymonta 4, 30-059 Krakow, Poland}
}
\author{Z. Burda}{
address={M.Smoluchowski Institute of Physics, Jagellonian University,
         Reymonta 4, 30-059 Krakow, Poland},
}

\author{B. Wac\l{}aw}{
address={M.Smoluchowski Institute of Physics, Jagellonian University,
         Reymonta 4, 30-059 Krakow, Poland}
}

\begin{abstract}
  Growing networks have a causal structure. We show that the causality
  strongly influences the scaling and geometrical properties of the
  network.  In particular the average distance between nodes is
  smaller for causal networks than for corresponding homogeneous
  networks. We explain the origin of this effect and illustrate it
  using as an example a solvable model of random trees.  We also
  discuss the issue of stability of the scale-free node degree
  distribution.  We show that a surplus of links may lead to the
  emergence of a singular node with the degree proportional to the
  total number of links.  This effect is closely related to the
  backgammon condensation known from the balls-in-boxes model.
\end{abstract}

\maketitle

\section{Introduction}

Many real-world complex networks have heavy tails in the node degree
distribution \cite{ref:ab,ref:dm,ref:n1}. 
It was a great discovery \cite{ref:bamodel} to realize that networks with
such a property can be naturally produced by a preferential attachment
growth \cite{ref:s}. It has become an
important issue to work out the consequences of the presence of heavy
tails in the degree distribution on topological properties of
networks \cite{ref:ab,ref:dm,ref:n1}. 

The degree distribution however gives only a crude information about
topology of random networks unless it is supplemented by some other
information.  Let us take as an example $k$-regular graphs. They are
defined as graphs with all nodes having degree equal $k$. The node
degree distribution of $k$-regular graphs is $p(q) = \delta_{qk}$
independently of the graph topology. A regular equilateral
triangulation or a cubic lattice are perfect examples of $6$-regular
graphs but they have completely different topological and geometrical
properties: the triangulation has fractal dimension two while the
cubic lattice -- three.  One can see the difference when one
determines for example the number of nodes in the given distance from
the given node.

The two examples of $6$-regular graphs are not random graphs so one
could argue that maybe if one considered random graphs from the
ensemble of $6$-regular graphs one would see statistically identical
topological and geometrical properties of them.  But to answer this
question one has to define what a random graph is. One can
do it by introducing a statistical ensemble of graphs
\cite{ref:bck,ref:k,ref:bl1,ref:dms,ref:bjk1,
ref:fdpv,ref:pn1,ref:lgkk,ref:bbw}.  
This means that one has to specify the set of graphs which one wants to study: as for instance
$k$-regular graphs, trees, simple graphs, pseudographs, etc.,
their attributes: directed, undirected,
Eulerian, connected, etc. Then one has to define a probability measure
on this set by ascribing to each graph from the set a positive number.
This number gives (after normalization) the probability that the graph
will be selected when the set is sampled randomly.  The statistical
properties of the ensemble heavily depend on the choice of the
probability measure.

The classical example of this approach is the ensemble of
Erd\"{o}s-R\'{e}nyi graphs \cite{ref:er1,ref:er2,ref:bb}. 
The set of graphs consists of all graphs 
with $N$ nodes and $L$ links
having neither multiple- nor self-connections.  The measure in the
Erd\"{o}s-R\'{e}nyi ensemble is defined as follows \cite{ref:bbw}.
One labels all nodes by the integers $1,\dots, N$ to obtain
labeled graphs. Such graphs are isomorphic to $N\times N$ symmetric
adjacency matrices with $L$ entries equal to one in the upper
triangle (and symmetrically also in the lower triangle).  Each labeled
graph (adjacency matrix) has the same statistical weight. The
partition function reads:
\begin{equation}
Z_h = \sum_{lg} 1 = \sum_{g} n_h(g).
\label{hg}
\end{equation}
The index $h$ stands for homogeneous and will be explained later.  The
first sum runs over all labeled graphs, the second over graphs
(=unlabeled graphs).  By a graph we mean graph's topology that is the
shape or skeleton which one obtains when one removes the labels. The
number $n_h(g)$ of distinct labelings (adjacency matrices) of the
graph $g$ depends on $g$ and thus some graphs are more whereas some
are less probable in this ensemble. 

One should realize that the definition \eqref{hg} assumes that a) we
can distinguish the vertices of the graph, b) we
are only interested in properties that do not depend on the labelings.
The permutation of indices neither changes graphs topology nor
physical quantities measured on this graph. The model has thus
permutation symmetry and one should divide out the volume of the
permutation group. One could explicitly write the factor of $1/N!$
instead of $1$ in the sum \eqref{hg}. However, as long as $N$ is fixed
this factor is constant and therefore can be pulled out in front of
the sum and skipped as an irrelevant normalization of the partition
function. One should keep it in any formula for an ensemble with
varying $N$ as we will do in the next section (see for example 
\eqref{th}). The origin of this factor is the same as in classical
statistical mechanics for identical particles.  These 
assumptions are reasonable for many real--world examples of graphs, but one must be
aware that they can be violated. For example when considering chemical
compounds we must treat the vertices of the same kind as indistinguishable
similarly as in quantum statistics. Then we should use an ensemble
where each graph topology has the same weight: $Z=\sum_{g} 1$.  In
practice such a definition turns out to be much more difficult to
handle \cite{ref:polya}. Also the ensembles, where the probability of
selecting a given graph depends on its labels, appear naturally when one
considers ensembles of growing
networks \cite{ref:bbjk-causal,ref:b-causal}.

In the Erd\"os-R\'enyi ensemble (\ref{hg}) all labeled graphs are
equiprobable. One can however weight the graphs in a way which depends
on their topology: for example one can introduce correlations between
neighboring nodes degrees \cite{ref:bl1,ref:b-correlations}, 
one can favor loops \cite{ref:bjk1}
which are very rare
for the Erd\"os-R\'enyi graphs, or one can introduce a weight which modifies
the node-degree distribution \cite{ref:bck,ref:bk}. 
In the last example one does it by
modifying the partition function to the following form:
\begin{equation}\label{weights}
Z_h = \sum_{lg} w(q_1) w(q_2) \dots w(q_N),
\end{equation}
where $q_k$ is the degree of $k$-th node and $w(q)$ is an
arbitrary weight function. One can tune the weight
function $w(q)$ in order to obtain graphs with a desired probability distribution.
In this way one can produce for example scale-free graphs with the
Barab\'{a}si-Albert degree distribution \cite{ref:bck,ref:bbw}.

In this contribution we apply the concepts of statistical
mechanics like statistical ensembles, partition function, averages over
ensemble, and so forth to study random graphs. 
Within this approach one has already obtained
many interesting results
\cite{ref:bck,ref:k,ref:bl1,ref:dms,ref:bjk1,ref:fdpv,ref:pn1,ref:lgkk,ref:bbw}.  
This approach is a straightforward
generalization of the Erd\"os-R\'enyi ideas.

At the first glance one can think that the statistical mechanics is not
adequate for growing networks which are not in equilibrium. Indeed such
ensembles cannot be understood as equilibrium
ensembles with the Gibbs measure, having temperature etc.  One should
rather understand them as follows. Imagine that one repeats the
process of growth many times independently and each time one
terminates it when the network reaches a given size.  One obtains a
collection of networks which occur with a certain probability.  This
perfectly defines an ensemble of graphs \cite{ref:k,ref:bbjk-causal}.  
It is not an equilibrium
ensemble but all methods of statistical mechanics work, so one can use
them. In order to distinguish growing networks
which are inhomogeneous from homogeneous graphs discussed above 
(for which all nodes are treated equally as for
instance for Erd\"os-R\'enyi graphs) 
we shall call the growing networks - {\em causal},
and the networks obtained from arbitrarily labeled graphs - {\em homogeneous}.
The difference will become clear in the next section.  Anticipating
some results, the causal and homogeneous networks have different
geometrical features even if they have identical node-degree
distributions.

\section{Causal versus homogeneous networks}

\begin{figure}
\psfrag{a1}[c][lb][1][0]{1}
\psfrag{a2}[c][lb][1][0]{2}
\psfrag{a3}[c][lb][1][0]{3}
\psfrag{a4}[c][lb][1][0]{4}

\psfrag{b1}[c][lb][1][0]{1}
\psfrag{b2}[c][lb][1][0]{3}
\psfrag{b3}[c][lb][1][0]{4}
\psfrag{b4}[c][lb][1][0]{2}

\psfrag{c1}[c][lb][1][0]{1}
\psfrag{c2}[c][lb][1][0]{2}
\psfrag{c3}[c][lb][1][0]{4}
\psfrag{c4}[c][lb][1][0]{3}

\includegraphics[width=14cm]{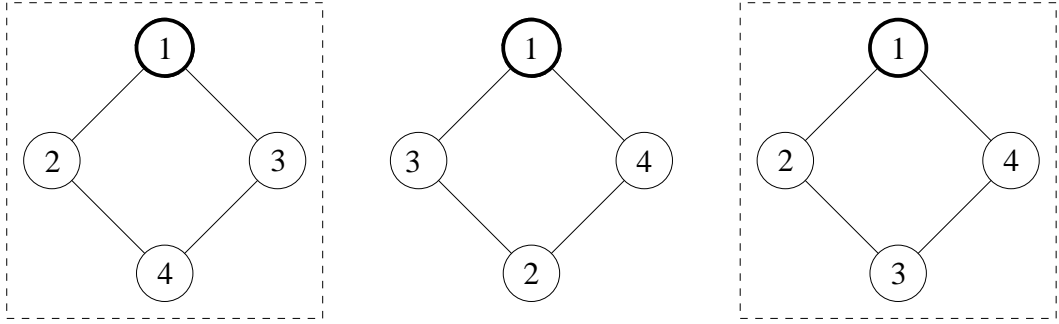}

\caption{\label{fig:causal-graph} Examples of labelings of a simple
  graph. Two labelings marked with the dashed rectangle are causal.
  The labeling in the middle is not. Also the remaining nine labelings
  of this graph (not shown on the picture) are not causal because the root
  does not have the label one. }
 \end{figure}

Vertices of a growing network can be labeled by integers representing
the order of attachment to the network.  It is clear that not all
possible labelings of the underlying graph can be realized in this way
(see figure \ref{fig:causal-graph}). We will call those that can be -- 
{\em causal} labelings or equivalently we will say that the labels are
causally ordered. A necessary condition for a causal labelings is that
a) root has the smallest label and b) one can connect every vertex to
the root\footnote{Obviously we are considering only connected graphs.
For disconnected graph this condition must be suitably modified.} by
a path (not necessarily the shortest) in such a way that the labels on
it are increasing from the root. It implies a condition
that every vertex except the root must have at least one
neighbor with a smaller label.

As in the
previous section we can define a partition function for the ensemble
of equiprobable rooted graphs with causal ordering:
\begin{equation}
Z_{cNL} = \sum_{clg} 1 = \sum_{g} n_c(g) .
\label{cg}
\end{equation}
The first sum is done over all causally labeled graphs $clg$.  This
ensemble is a counterpart of the Erd\"os-R\'enyi ensemble \eqref{th}
for homogeneous graphs in the sense that all labeled graphs are
equiprobable.  The difference between the two is that only causal
labelings are allowed here.  The second sum in (\ref{cg}) is done
over all graphs $g$. Each graph is weighted in the sum by the number
of its causal labelings.  This number is smaller than the number of
all labelings $n_h(g)$ of this graph. Although the partition
functions (\ref{hg}) and (\ref{cg}) cover the same set of graph
topologies $\{g\}$, their statistical weights are different.  When one
samples graphs randomly in the homogeneous graphs ensemble (\ref{hg})
one thus observes different probabilities of graphs' (topologies') occurrence  
than in the ensemble of causal graphs.

This can be illustrated by calculating various quantities for random
tree graphs (branched polymers) for which the calculations can be done
analytically \cite{ref:bbjk-causal,ref:b-causal,ref:b-correlations,ref:adfo,ref:adj,ref:bb-trees,
ref:jk-polimers,ref:bbj,ref:bepw}.

\begin{figure}
\psfrag{=}{$=$}
\psfrag{+}{$+$}
\psfrag{1}{$ $}
\psfrag{2}{$\frac{1}{1!}$}
\psfrag{3}{$\frac{1}{2!}$}
\psfrag{d}{$\cdots$}
\includegraphics[height=.15\textheight]{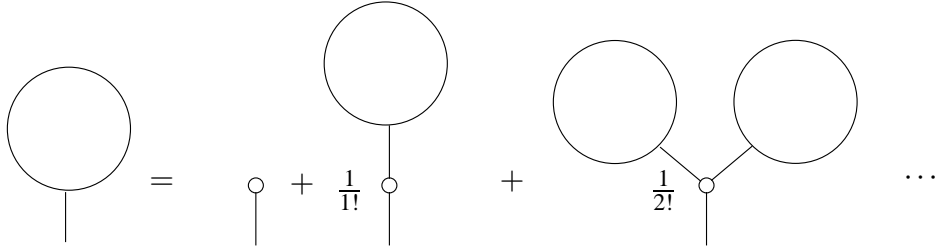}
\caption{\label{fig1} Graphical representation
of the self-consistency equation \eqref{z_h}.
The bubble contains the grand-canonical sum over
planted rooted trees. Each tree is weighed with the size 
factor $x^N$ and the small circle representing 
a single node gives additional factor $x$.}
\end{figure}

Let us give a short account on results of those calculations.  We shall
discuss ensembles of planted rooted trees.  Each graph in such an
ensemble is connected and has no loops.  One node of the graph has a
single line sticking from it interpreted as the stem of the tree (we
omit the root node at the end of the stem).  The
number of branches (links) $L$ of a tree with $N$ nodes is equal to
$N-1$. The stem is not counted as a branch.  Because $L$ is related to
$N$ we can skip $L$ in the $Z_{hNL}$ and denote the canonical partition
function for trees by $Z_{hN}$.  The grand canonical
partition function for homogeneous planted trees is
\begin{equation}
Z_h(x) \equiv \sum_{N=1}^\infty \frac{Z_{hN}}{N!} x^N
\label{th}
\end{equation}
and can be represented as a bubble in figure \ref{fig1}. The free
end of the bubble denotes the stem while the
bubble -- the sum over all trees weighted as in \eqref{th}.
One can write an identical definition for causal trees.
For technical reasons instead of the chemical potential $\mu$ 
in the grand-canonical partition function \eqref{th}
we prefer to use the fugacity $x = e^{-\mu}$.

An advantage of using the grand-canonical partition function
is that one can deduce a closed formula \cite{ref:bbjk-causal} expressing
$Z_h(x)$ as a function of itself:
\begin{equation}
Z_h(x) = x \sum_{k=0}^{\infty} \frac{1}{k!}Z_{h}^{k}(x)= x \ e^{Z_h(x)},
\label{z_h}
\end{equation} 
which can be graphically represented as in figure \ref{fig1}.  The
meaning of this graphical equation is the following.  Any rooted tree
can be constructed by joining stems of a certain number of trees at a
common node and attaching a new common stem to this node.  If one sums
over all trees in each bubble on the right-hand side of the graphical
equation shown in figure \ref{fig1}, one obtains a sum over all trees
also on the left-hand side.  In order to obtain the partition function
$Z_h(x)$ \eqref{th} one has to take care of the power of $x$ which
counts the number of vertices. We see that the new tree has by one
vertex more than all subtrees used in the construction so one has to
multiply the right-hand side by $x$.  In order to avoid overcounting
while joining $k$ trees, which now form $k$ branches of the new tree,
one has to divide out the factor $k!$ which counts the number of
indistinguishable ways in which one can put the trees together. Thus
the expression $Z_h^k$ which represents the composition of $k$
subtrees is divided by $k!$. Finally, one has to sum over $k$ to
include all branching possibilities at the root. 
In this way one reproduces the partition
function $Z_h$ on the left-hand side and $x \exp (Z_h)$ on the
right-hand side of Eq.~\eqref{z_h}.

\begin{figure}
\includegraphics[height=.3\textheight]{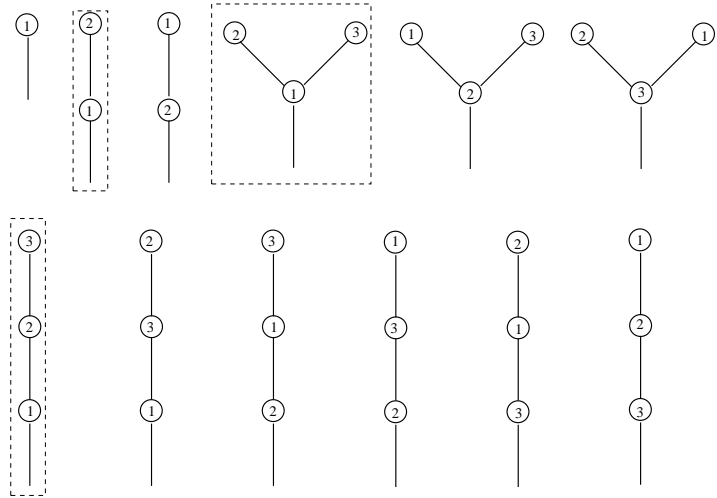}
\caption{\label{fig2} Planted rooted trees of
size $N=1,2,3$. Causal labelings are marked with dashed rectangles.}
\end{figure}

Using the equation \eqref{z_h} one can determine $Z_{hN}$ by inverse Laplace transform (here written as
a contour integral):
\begin{equation} 
Z_{hN} = \frac{N!}{2\pi i} \oint \frac{{\rm d} x}{x^{N+1}}  Z_h(x).
\end{equation}
Changing the integration variable from $x$ to $Z=Z_h(x)$ and by means of
\eqref{z_h} one obtains
\begin{equation}
Z_{hN} = \frac{N!}{2\pi i} 
\oint \frac{{\rm d} Z}{Z^N} e^{NZ} \left(1-Z\right)
= N! \left(\frac{N^{N-1}}{(N-1)!} - \frac{N^{N-2}}{(N-2)!}\right) = N^{N-1}.
\label{rooted}
\end{equation} 
This result tells us that there are $N^{N-1}$ labeled planted rooted trees
of size $N$. There is one planted rooted tree for $N=1$, 
two for $N=2$, nine for $N=3$, etc. The labeled trees for $N=1,2,3$
are shown in figure \ref{fig2}.
Note that if one removes the stem one obtains an ensemble with one
marked vertex - the tree is not planted anymore. The number of trees
with one marked vertex is thus also equal to $N^{N-1}$. From this one can
deduce the number of all labeled trees without any marked vertex.  On
a tree with $N$ vertices one can mark one of $N$ vertices, so the
number of trees with a marked vertex must be $N$ times larger than the
total number of labeled trees. Thus the number of labeled trees of size $N$ is
$N^{N-2}$.  This is the classical result derived first by Cayley (see
reference \cite{ref:enum} for general introduction on counting
graphs).

Let us come back to planted rooted trees, but now consider only
causally labeled ones.  For the trees we have exactly one path joining
any given vertex and the root and along this path labels must increase from
the root outwards (see figure~\ref{fig2}). One can derive
\cite{ref:bbjk-causal} a self-consistency equation for causal planted
rooted trees similarly to \eqref{z_h}.  It has the same logical
structure as the graphical equation shown in figure \ref{fig1} but the
causality imposes a requirement on labels ordering. A careful analysis
of the consequences implied by this requirement shows that one obtains
the following self-consistency equation for causal trees 
\cite{ref:bbjk-causal}:
\begin{equation}
\frac{\mbox{d}Z_c(x)}{{\rm d}x} =  e^{Z_c(x)}.
\label{z_c}
\end{equation}
This equation can be solved for $Z_c$ yielding:
\begin{equation}
Z_c(x) = - \ln (1 - x) = \sum_{N=1}^\infty \frac{1}{N} x^N
 = \sum_{N=1}^\infty \frac{Z_{cN}}{N!} x^N,
\end{equation}
and hence:
\begin{equation}
Z_{cN} = (N-1)!.
\end{equation}
Thus there are $(N-1)!$ causally labeled planted rooted trees.  For
example, there is one for $N=1$, one for $N=2$ and two for $N=3$, etc.
One can easily find these trees among all labeled trees in figure
\ref{fig2}.  Causally labeled trees form a small subset of all labeled
trees: the fraction of causally labeled trees $Z_{cN}/Z_{hN} \sim
N^{3/2} e^{-N}$ quickly disappears when the size of the system grows.
One may ask whether the statistical properties of this subset are
identical as of the whole set. The answer is that they are completely
different. Probably the most striking difference is that the causal
trees are much more compact and have much smaller linear extent than
homogeneous trees.

One can quantify this statement by checking how the average distance
between nodes depends on the size of the tree.  One defines
geodesic distance $r_{ab}$ for each pair of vertices $a,b$ as the
number of links of the (shortest) path between $a$ and $b$. To get the
average distance one averages $r_{ab}$ over all pairs of nodes on the
tree and over all trees in the ensemble. The result is that for
homogeneous trees the average distance behaves for large $N$ as
\cite{ref:b-correlations,ref:adj,ref:bbj}:
\begin{equation}
\langle r \rangle_h \sim \sqrt{N},
\label{rh}
\end{equation}
while for causal 
ones\footnote{In the references \cite{ref:bbjk-causal,ref:kr-growing} 
only the distribution of distances from the root was calculated.} 
\cite{ref:bbjk-causal,ref:kr-growing}:
\begin{equation} 
\langle r \rangle_c \sim \ln {N}.
\label{rc}
\end{equation}
The power-law growth of the linear extent $\langle r \rangle_h \sim
N^{1/2}$ with the system size means that homogeneous trees have
fractal dimension $d_h=2$, while the behavior $\langle r \rangle_c
\sim \ln N$ that the fractal dimension of causal trees is
$d_c=\infty$.  The separation of vertices on a causal tree is very
small and the whole tree structure is compactly concentrated around
the oldest part of the tree.

\begin{figure}
\psfrag{xx}{$r$}
\psfrag{yy}{$G_N^{(2)}(r)$}
\includegraphics[height=.3\textheight]{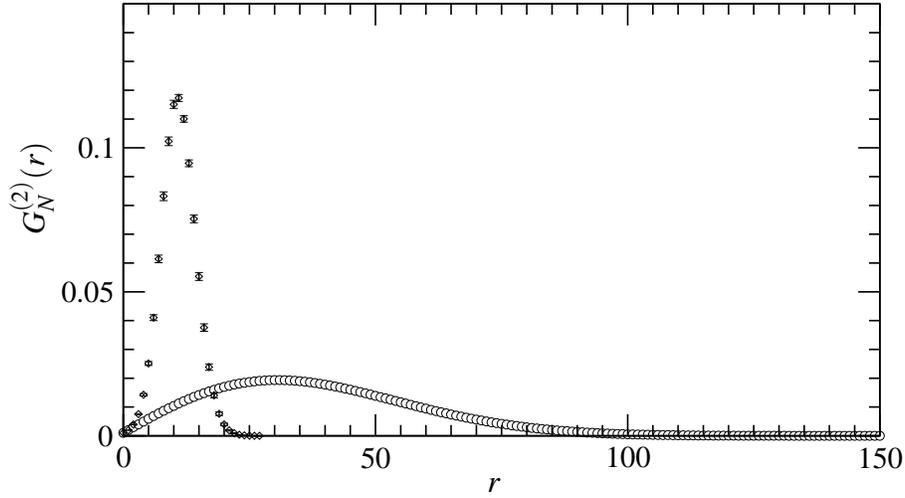}
\caption{\label{fig3} The distance distribution functions
for the homogeneous (circles) and causal trees (diamonds) for $N=1000$.
Error bars for data points represented by circles are smaller than the 
symbol size.}
\end{figure}

This can be  well illustrated by studying the two-point
correlation function $G_N^{(2)}(r)$:
\begin{equation}
G_N^{(2)}(r) = \left\langle \frac{1}{N^2} \sum_{ab} \delta(r-r_{ab}) 
\right\rangle .
\end{equation} 
The delta function $\delta(r-r_{ab})$ selects only pairs
$a,b$ separated by $r$ edges.
The two-point function can be thus interpreted as the probability
that two randomly chosen nodes of the tree are separated by
$r$ links. Alternatively it can be interpreted as the distance
distribution giving the fraction
of nodes at a distance $r$ from a randomly chosen one.
By construction, $\sum_r G_N^{(2)}(r) = 1$ and $G_N^{(2)}(0) = 1/N$.
The average distance between vertices is given by the
mean value of the two-point function:
\begin{equation} 
\langle r \rangle = \sum_{r=1}^\infty r \; G_N^{(2)}(r) .
\end{equation}
The two-point function gives a very valuable information about 
the distance distribution in the given ensemble of graphs. 
In figure \ref{fig3} we compare 
the two-point function $G_{hN}^{(2)}(r)$ for homogeneous and
$G_{cN}^{(2)}(r)$ for causal trees of the same size $N=1000$.
One sees that indeed the typical causal trees are much
shorter than homogeneous. The distance distribution 
for causal trees is much more concentrated.

One can analytically derive the shape of the two-point function
$G_{hN}^{(2)}(r)$ for homogeneous trees in the limit of $N\rightarrow
\infty$. In this limit one can approximate the shape by a function of
a continuous variable:
\begin{equation}
G_{hN}^{(2)}(r) \approx \frac{ar}{N} \exp\left( -\frac{ar^2}{2N} \right),
\label{ghN}
\end{equation}
where $a$ is a positive constant. We have:
\begin{equation}
\int_0^\infty \kern-0.5em {\rm d} r \ G_{hN}^{(2)}(r) = 1 \ , 
\end{equation}
and
\begin{equation}
\langle r \rangle_h \approx \int_0^\infty \kern-0.5em{\rm d} r 
\ r G_{hN}^{(2)}(r) = \frac{\sqrt{N}}{\sqrt{a}} .
\end{equation}
In the large $N$ limit the two-point function
becomes effectively an $N$-independent function of 
the scaling variable $x = r/\sqrt{N/a}$:
\begin{equation}
G_{hN}^{(2)}(r) {\rm d} r = g_h(x) {\rm d} x = x e^{-x^2/2} {\rm d} x.
\label{sghN}
\end{equation}
When one draws the two-point functions for different system sizes as a
function of the scaling variable the corresponding curves collapse to
a universal shape independent of $N$.  It will be illustrated below
for the scale-free graphs.

The situation is completely different for causal (growing) trees 
\cite{ref:sd-pk}. First of all, the mean distance 
$\left<r\right>_c$ grows like $\ln N$ and not like $\sqrt{N}$.
Second of all, the two-point
functions $G^{(2)}_{cN}(r)$ for different sizes do not collapse after the rescaling $x=r/\ln N$. 
If the scaling hold, the dispersion of $G^{(2)}_{cN}(r)$
would grow like $(\ln N)^2$. We shall see below that it grows only like $\ln N$.
Let us first calculate the average number of vertices 
of the causal tree $n(r,N)$ in the distance $r$ from the root - 
a quantity which is closely related to the two point function.
Since the causal tree is constructed recursively by adding 
new vertices we can write the following recursive relation:
\begin{equation}
	n(r,N+1) = n(r,N)+n(r-1,N)/N,	\label{eq:nrec}
\end{equation}
which tells us that the new vertex added
in the distance $r$ from the root must be linked
to a vertex in the distance $r-1$. The factor $1/N$
is just the probability of choosing one out of $N$
vertices on the tree, and $n(r-1,N)/N$ is the probability
that we choose a vertex in the distance $r-1$ from the root.
Dividing $n(r,N)$ by $N$ we obtain the probability that a randomly chosen node is at distance $r$ to the root:
\begin{equation}
	G_{\rm root}(r,N) \equiv n(r,N)/N.	\label{eq:}
\end{equation}
Multiplying Eq.~(\ref{eq:nrec}) by $r^m$ and summing over $r$ we can calculate
moments $\left<r^m\right>_{\rm root}$ of this probability distribution.
In particular, the first two moments read:
\begin{eqnarray}
	\left<r\right>_{\rm root} = -\frac{1}{2} + H(N), \label{eq:rHn} \\
	\left<r^2\right>_{\rm root} = \sum_{n=2}^N \frac{2}{n} H(n-1),
\end{eqnarray}
where $H(n) = \sum_{i=1}^n 1/i$ is the harmonic number. 
Because $H(N)\cong \ln N$ for large $N$, 
the mean distance $\left<r\right>_{\rm root}$ 
behaves as $\ln N$. But the dispersion 
$\sigma^2_{\rm root}(r) = \left<r^2\right>_{\rm root} -
\left<r\right>_{\rm root}^2 
\cong \ln N$ grows also logarithmically. So if one introduced
a scaling variable $x=r/\ln N$, one would have in the limit 
$N\rightarrow \infty$: $\left<x\right>_{\rm root} \rightarrow \mbox{const}$ 
and $\sigma^2_{\rm root}(x) \sim \mbox{const}/\ln N \rightarrow 0$,
so it is not a good scaling variable.
Actually, one can solve the recursion relation (\ref{eq:nrec}) using a generating function formalism.
One obtains in the limit $N\rightarrow \infty$:
\begin{equation}
	G_{\rm root}(r,N) \approx \frac{1}{\sqrt{2\pi\sigma^2}} \exp\left[-\frac{(r-\left<r\right>)^2}{2\sigma^2}\right],
\end{equation}
where, as mentioned before, both $\left<r\right>$ and $\sigma^2$ grow logarithmically. 

The function $G_{\rm root}(r,N)$ is closely related to $G_{cN}^{(2)}(r)$,
the two-point function we introduced before.
For large growing trees, $G_{\rm root}(r,N)$ becomes almost identical to $G_{cN}^{(2)}(r)$,
because in almost all cases if one randomly chooses a pair 
of vertices, the shortest path between them goes through the root. 
Only if the two lie on the same branch, the
shortest path does not contain the root, but the probability that
they are on the same branch vanishes for $N\to\infty$. Thus for large
tree the difference between $G_{cN}^{(2)}(r)$ and $G_{\rm root}(r,N)$
is in the proportionality factor in the average distance:
$\left<r\right>_c \approx 2 \left< r\right>_{\rm root} \sim 2 \ln N$. To summarize,
the two point function behaves in a completely different way for
causal and homogenous trees.

Also the other statistical characteristics
of the causal trees are different from those for the homogeneous
trees. For instance one can determine the node degree distribution
averaged over trees in the ensemble:
\begin{equation}
p(q) = \left\langle \frac{1}{N} \sum_{a} \delta(q-q_a) \right\rangle .
\end{equation}
Here $q_a$ is the degree of node $a$, the sum runs over all
nodes of the tree, and the average is over all trees in the
ensemble. If one calculates this distribution
for the homogeneous trees one obtains:
\begin{equation}
p_h(q) = \frac{e^{-1}}{(q-1)!},
\end{equation}
while for causal trees:
\begin{equation}
p_c(q) = 2^{-q} .
\end{equation}
Again we see the difference between the set of all labeled
trees and the subset of causally labeled trees.

So far we have discussed the ensembles of unweighted graphs: each labeled
graph has had a statistical weight equal to $1/N!$ independently of
its shape. One can easily extend those considerations to the ensembles
of weighted graphs, in particular to an ensemble of graphs
for which the statistical weight
includes a factor depending on node degrees \eqref{weights}.  For
instance, one can analytically solve a model of trees (both causal and
homogeneous) whose statistical weight is a product of
weights depending on individual node degrees. For each node one
introduces a weight which is a function of the number of
edges emerging from it.
By tuning the weights one can obtain trees with a desired degree
distribution. However, the following question arises: assume that we
have two ensembles of trees with the same node degree distribution. In
the first ensemble we have all labeled trees and in the second only
causally labeled ones.  We now remove labels from the trees. All we
have is the graph topology. Can we distinguish whether a given graph
topology comes from homogeneous or causal graph ensemble?  One can answer
this question in the affirmative.

\begin{figure}
\psfrag{xx}{$q$}
\psfrag{yy}{$p(q)$}
\includegraphics[height=.3\textheight]{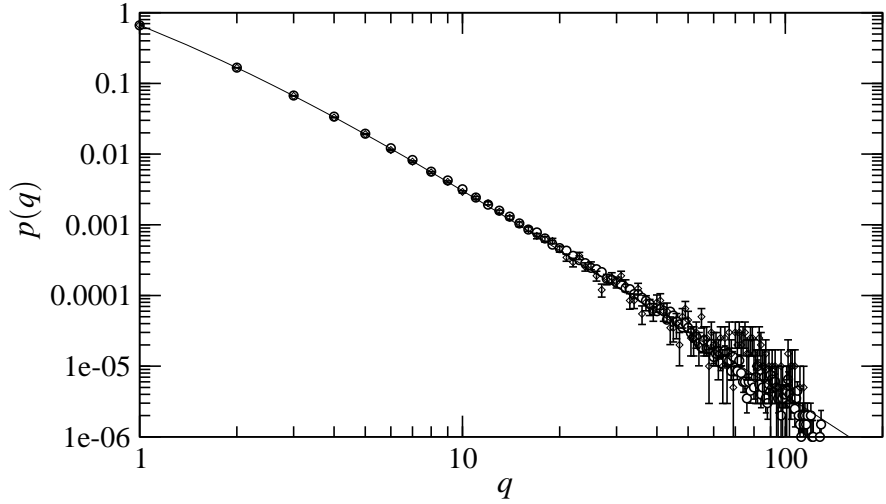}
\caption{\label{fig4} The degree distribution for causal (diamonds) and homogeneous (circles) scale-free trees
measured in Monte-Carlo runs for $N=1000$. One sees that they are statistically
identical.}
\end{figure}

\begin{figure}
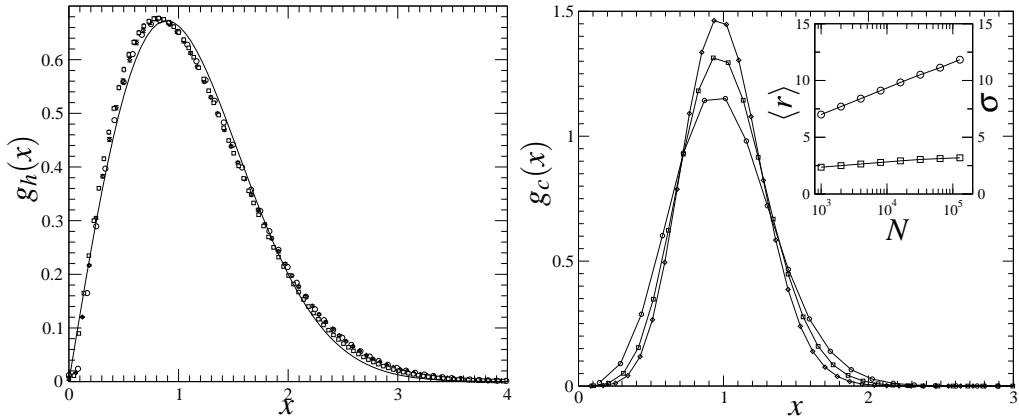

\psfrag{xx}{$x$}
\psfrag{yy}{$g_h(x)$}
\includegraphics[height=.25\textheight]{fig5a.eps}
\hspace{1mm}
\psfrag{yy}{$g_c(x)$}\psfrag{rr}{$\left<r\right>$}\psfrag{ss}{$\sigma$}\psfrag{nn}{$N$}
\includegraphics*[height=.25\textheight]{fig6b.eps}
\caption{\label{fig5} Left: The distance distributions for 
homogeneous trees with BA degree distribution (\ref{ba}) plotted in the rescaled variable 
$x=r/\sqrt{N/\ln N}$ for different 
$N$: $N=500,1000,2000,4000$. 
The continuous line is given by the function $b x \exp (-ax^2/2)$.
Right: The distance distributions for 
causal trees with the same distribution (\ref{ba}), plotted in the rescaled variable 
$x=r/\ln N$ for three different sizes $N=1000,16000,128000$.
The inset shows how the average distance $\left<r\right>$ and the width $\sigma$ of $G^{(2)}_{cN}(r)$ scale with the system size,
for $N=1000,2000,\dots,128000$. Both the plot and the inset indicate that $\sigma$ grows slower than $\left<r\right>$ and thus the curves $g_c(x)$ become
more narrow while $N$ increases. To observe this effect one has to go to much larger sizes than for homogeneous trees.
}
\end{figure}

Let us sketch the solution. We will generate scale-free tree graphs by
tuning the nodes weights. Introducing this to
the model, one has to modify the right-hand side of the
self-consistency equations. The exponential series which we used
before in equations \eqref{z_h} and \eqref{z_c} to generate equally
weighted trees has to be substituted by:
\begin{equation}\label{series}
\sum_{k=0}^\infty \frac{Z^k}{k!} \rightarrow 
\sum_{k=0}^\infty \frac{w(k+1) Z^k}{k!} .
\end{equation}
Here $w(q)$ is the node weight which depend only
on node's degree $q$. It can be also interpreted as branching
ratio that a link will split into $q-1$ links. 
In the previous case the weights $w(q)=1$ were identical for all $q$'s.
In order to produce scale-free networks  
with the Barab\'{a}si-Albert distribution:
\begin{equation}
p_{BA}(q) = \frac{4}{q(q+1)(q+2)},
\label{ba}
\end{equation}
one has to choose $
w_{c}(q) = (q-1)! $ for causal trees and $w_{h}(q) =
(q-1)!p_{BA}(q)$ for homogeneous trees.  In the limit of
$N\rightarrow\infty$ these two ensembles have
identical node-degree distributions $p_c(q)=p_h(q)=p_{AB}(q)$. To
illustrate this we show in figure \ref{fig4} node degree distributions obtained by
Monte-Carlo simulations of causal and homogeneous trees with the
appropriately chosen branching weights. One can see that
they perfectly follow the distribution \eqref{ba}. In other words, one cannot
distinguish to which ensemble the tree belongs just by measuring nodes
degrees.
One can however distinguish the ensembles very easily if one determines the
distance distribution $G^{(2)}_N(r)$.  As before, the causal trees have
an infinite fractal dimension, while the homogeneous ones the fractal
dimension equal to two. 
The distance distribution $G^{(2)}_N(r)$
for homogeneous trees is plotted 
in the scaling variable $x=r/\sqrt{N/\ln N}$
on the left-hand side of figure \ref{fig5}
and for causal trees in the scaling variable $x=r/\ln N$
on the right-hand side. 
As expected, the data for homogeneous trees collapse to a curve independent of
$N$. In the case of causal trees there is no collapse due to a weak dependence of the curve's width
$\sim 1/\sqrt{\ln N}$ on the size $N$, similarly to the case of unweighted trees mentioned before.
The average distance scales as 
$\langle r \rangle_h \sim \sqrt{N/\ln N}$
in the first case while as $\langle r \rangle_c \sim \ln N$
in the second. Note that we have introduced a logarithmic
correction to the square root scaling \eqref{rh} and \eqref{sghN} for
the scale-free homogeneous trees.  This is due to the fact that in
this case the series \eqref{series} develops a logarithmic singularity. 
This correction does not affect the
fractal dimension which is still two.  The average distance scales as
\begin{equation}
\langle r \rangle_h \sim \sqrt{N/\ln N} \ .
\label{log_corr}
\end{equation}
The distribution \eqref{ba} belongs to a broader class of 
scale-free distributions \cite{ref:kr-growing}:
\begin{equation}
p_{KR}(q) = \frac{(2+\omega)\Gamma(3+2\omega)}{\Gamma(1+\omega)}
\cdot \frac{\Gamma(k+\omega)}{\Gamma(k+3+2\omega)} ,
\label{kr}
\end{equation}
which emerge as the limiting distributions in a growth process 
with the linear attachment kernel $A_q = q + \omega$, $\omega > -1$.
The BA distribution \eqref{ba} corresponds to $\omega=0$.
The mean value of the $p_{KR}(q)$ distribution is
equal two: $\sum_q q p_{KR}(q) = 2$ for all $\omega>-1$, in 
accordance with the average number of links per node on a tree.
For large $q$ the distributions have a power-law tail:
$p_{KR}(q) \sim q^{-\gamma}$
with the exponent $\gamma = 3 + \omega$, which assumes
the value from the range $\gamma>2$. Interestingly enough, 
for the scale-free causal trees \eqref{kr} the average
distance scales logarithmically $\langle r \rangle_c \sim \ln N$
independently of $\gamma$, whereas the scaling properties
of the homogeneous trees strongly depend on $\gamma$. In particular
the average distance scales as $\langle r \rangle_c \sim N^{1/d_f}$
where \cite{ref:bb-trees,ref:jk-polimers,ref:bbj} 
\begin{equation}
d_f = \mbox{max}\left(2,\frac{\gamma-1}{\gamma-2}\right) .
\label{df}
\end{equation}
We have $d_f=2$ for $\gamma > 3$.  For $2<\gamma <3$ the fractal
dimension $d_f$ assumes values which continuously grow from two to
infinity while $\gamma$ decreases from three to two.  Please note that
the logarithmic corrections appear for $\gamma=3$. Those corrections can
be interpreted as the fact that at this point $\langle r
\rangle$ grows slower than $\sqrt{N}$ but faster than  any
 power  $N^{1/2-\epsilon}$ with $\epsilon >0$. 

\begin{figure}
\psfrag{xx}{$r$}
\psfrag{yy}{$G^{(2)}_N(r)$}
\includegraphics[height=.3\textheight]{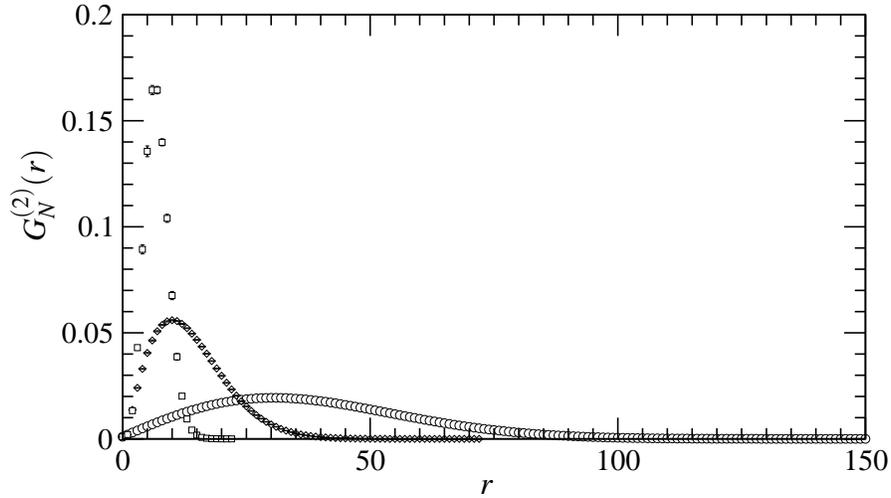}
\caption{\label{fig6} The distance distribution $G^{(2)}_N(r)$
for $N=1000$ for unweighted homogeneous tree (circles), scale-free
homogeneous trees (diamonds) and scale-free causal trees (squares).}
\end{figure}

The discussion of this section can be summarized as follows. We have
considered four ensembles of trees: (a) homogeneous uniformly
weighted, (b) causal uniformly weighted, (c) homogeneous with the
BA scale-free distribution \eqref{ba}, (d) causal weighted
with the BA scale-free distribution.  The average distance
between nodes in a scale-free system is generally smaller than in a
random one: so graphs in the ensemble (c) have on average smaller diameter
 than in (a) and in (b) than in (d). This effect is well
known.  It is related to the presence of nodes with high degree which
cluster around themselves many vertices just in distance one.  Another
effect which is less known is that the causality lowers the distances
between nodes on the graph. The effect caused by causality is even
stronger than by scale-free tails: the graphs in (b) have diameter
much smaller than those in (a), and similarly those in
(c) than those in (d).  One observes big changes if one imposes the
causality constraint.  The reason why causality plays such an important
role in enhancing the small world effect is related to the fact that
the oldest vertices, in addition to having highest degrees, cluster
with each other forming a kernel of the graph with extremely high
connectivity.  Remaining nodes tidily surround this compact kernel
making the whole graph structure jampacked.  In figure \ref{fig6} we
compare the distance distribution $G^{(2)}_N(r)$ for the cases (a), (c)
and (d) for the same system size. We see that homogeneous graphs (a)
are more elongated than causal ones (c) which in turn are more
elongated than causal scale-free ones (d).  We have not shown the case
(b) in the figure in order to keep it transparent. The case
(b) was compared to (a) in the figure \ref{fig3}.

\section{Condensation}

Statistical ensemble of causal trees discussed in the previous section
is for the weights $w(q) = \Gamma(q+\omega)/\Gamma(1+\omega)$
\cite{ref:bbjk-causal} 
equivalent to an ensemble of trees obtained by a growth process
with the linear attachment kernel $A_q = q + \omega$ \cite{ref:kr-growing}.
The map between these models is
mathematically exact.  For non-linear kernels the two models slightly
differ but both display the same features\footnote{see however the
  reference \cite{ref:b-causal}.}. If one applies a superlinear
attachment kernel in the growth process: $A_q \sim q^\sigma$ for
$\sigma>1$ one sees the appearance of a singular node which has the
degree proportional to the size of the tree. If one applies a
sublinear kernel $\sigma<1$, the degree distribution will be
suppressed exponentially for large $q$. The linear kernel is marginal
in the sense that it lies exactly between the phase with the singular
node and the exponential tail. A very similar situation takes place
for homogeneous trees but the mechanism of the emergence of the singular node
is different.  The model can be mapped onto a balls-in-boxes model
\cite{ref:b-i-b}.  In order to obtain scale-free trees one has to
choose appropriate vertex weights of the model. Any deviation
from the fine-tuned values results in either the appearance of the
singular node or the exponential suppression in the node
degree distribution, exactly as for growing networks. Roughly
speaking, if we denote the fine-tuned scale-free distribution $p_0(q)
\sim q^{-\gamma}$ (the second case in the equation below) we have
three possible scenarios:
\begin{equation}
p(q) \sim \left\{ \begin{array}{l} p_0(q) e^{-\mu q} \ ,\\
p_0(q) \ , \\
p_0(q) + \frac{1}{N} \delta( q - \rho N)  \ .
\end{array} \right.
\end{equation}
In the first case the typical fluctuations of the node degree are of
order $1/\mu$ independently of $N$.  In the third case there is a
singular node with an extensive number of links $q \sim \rho N$.  This is
equivalent to the backgammon condensation of the balls-in-boxes model
\cite{ref:b-i-b}.  The appearance of the singular node makes the
system to be even more compact than for scale-free graphs. The
distance between nodes increases slower than logarithmically since
many vertices are in the closest neighborhood of the singular node. In
the extreme case which corresponds to the star topology -- a vertex
surrounded by $N-1$ vertices -- the average distance is smaller than
two and is independent of $N$.

\begin{figure}
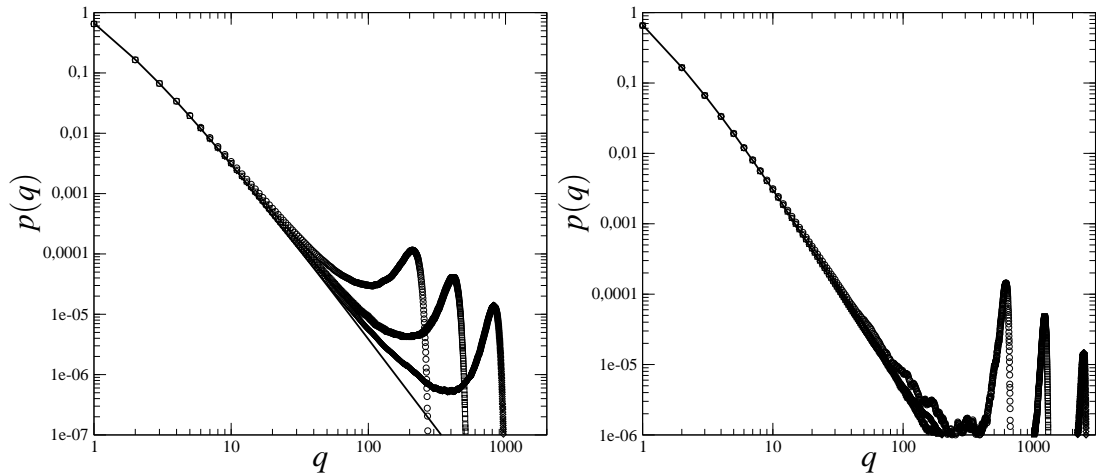

\psfrag{xx}{$q$}
\psfrag{yy}{$p(q)$}
\includegraphics[height=.28\textheight]{fig8a.eps}
\hspace{1mm}
\includegraphics[height=.28\textheight]{fig8b.eps}
\caption{\label{fig8} Left: The node degree distribution 
for scale-free pseudographs with $\bar{q}=2L/N = 4$ which is
above the condensation threshold $\langle q \rangle =2$ of the BA distribution \eqref{ba}.
The main part of the distribution 
goes along the limiting BA curve (solid line). The peak represents the
singular vertex. Its position moves linearly 
with the system size $N=200,400,800$.
Right: The same for larger link density:
$\bar{q}=2L/N = 8$.}
\end{figure}

Is the condensation a feature of the tree graph ensemble, or it is
observed for ensembles of graphs as well?  We have studied this
question for homogeneous scale-free graphs and pseudographs. In our
conventions graphs do not have multiple- and self-connecting edges
while pseudographs do. For graphs, the condition that they have
neither multiple- nor self-connections acts as strong constraints on
the graph structure which are sometimes called structural constraints
\cite{ref:bk,ref:bpv}.  In particular they strongly prevent the system
from developing a power-law tail in the node-degree distribution for
finite size systems \cite{ref:bk,ref:bpv}. 
It also turns out they prohibit the
condensation which we discussed above.  So far we have not found any
evidence for the backgammon condensation for graphs and the emergence of
singular node on the graph. On the contrary, for pseudographs the
situation is very much like for trees and one observes the
condensation. We simulated a canonical ensemble of homogeneous
pseudographs with $L$ links and $N$ vertices, with the distribution
$p_{BA}(q)$ \eqref{ba}. This was achieved by tuning the node degree
weights, similarly as we have described for trees before, and by
adjusting the ratio $\bar{q} = 2L/N$ by choosing $L=N$ to the mean
value of the distribution $\langle q \rangle = \sum_q q p_{BA}(q)
=2$.  Indeed the system produced the desired distribution.  The next
step was to check how the system reacts on the change of the number of
links. We had increased the number of links $L$ while keeping the number
of nodes $N$ constant so that the ratio $\bar{q}=2L/N=4$ exceeded
$\langle q \rangle=2$. The system reacted as follows.  The
distribution of the bulk part was as before equal
to the desired power-law distribution $p_{BA}(q)$ but
the system additionally produced a singular node 
which took the surplus of links.
The presence of the singular node is manifested as a peak in the
distribution. The position of the peak moves linearly with the system
size and the peak departs from the main body of the distribution.  Its
height is proportional to $1/N$ since it is a probability of picking
up one out of $N$ vertices. The situation is depicted in figure
\ref{fig8}. This is an example of the backgammon 
condensation \cite{ref:b-i-b}.
If one adds more links, then the surplus will go to the singular node.
One can see this effect by comparison of the plots in the left- and right-hand side of 
figure \ref{fig8} which differ by the ratio $\bar{q}=2L/N$
and the positions of the peak correspondingly.
If one chooses the ratio $\bar{q}=2L/N$ to be smaller
than $\langle q \rangle$ the system generates
an exponential tail in the node degree distribution. 
This is again what one expects from the balls-in-boxes model. The reason why
this analogy works for pseudographs and does not for graphs
is that for pseudographs the degrees of individual
vertices are almost independent of each other except of the
global constraint $q_1+q_2+\dots+q_N=2L$, while they
are in a subtle way correlated for graphs due to the
structural constraints.

\section{Summary}
We have applied methods of statistical
mechanics to compare ensembles of homogeneous
and causal (growing) networks. We have shown
that the causality strengthens the small world
effect due to clustering of 
nodes with high connectivity in the kernel
of the graph. In particular, for
homogeneous random trees the average inter-node
distance scales as $\sqrt{N}$ while for causal
networks as $\ln N$. We have compared two
ensembles of random trees with Barab\'{a}si-Albert
distribution and observed that despite they
have identical degree distribution, the causal
trees have much smaller diameter than the
corresponding homogeneous trees. 

We have also discussed the stability of the scale-free 
distributions. For growing network such
distributions emerge for linear attachments kernels.
If the attachment is slightly perturbed 
the system either exponentially suppresses nodes with
higher degree or develops a singular node with a degree
proportional to the total number of links. The first
type of perturbation corresponds to sublinear while the second
to superlinear kernels. Similar instabilities
are observed for homogeneous pseudographs. Scale-free
distributions require a fine-tuning of the weight parameters.
A slight perturbation, as before, also leads either 
to the exponential suppression
or to the emergence of a singular node on the graph.
In this case the singular node emerges as a result of
a condensation of the backgammon type \cite{ref:b-i-b}.

\begin{theacknowledgments}
This is essentially a review paper. Most of the results presented
above were obtained by the Krakow-Orsay collaboration, including Jerzy
Jurkiewicz and Andr\'{e} Krzywicki. We wish to thank Jerzy and 
Andr\'{e} for their contribution. 
We also thank S. N. Dorogovtsev and A. Krapivsky for the discussion
of the two-point function for causal trees.
ZB thanks the organizers of the conference CNET2004 for the
invitation. This work was partially supported by 
the Polish State Committee for Scientific Research (KBN) grant
2P03B-08225 (2003-2006) and by EU IST Center of Excellence ``COPIRA''.
\end{theacknowledgments}

\end{document}